\documentclass[prd,aps]{revtex4} 
\usepackage{graphicx}

\begin{document}
\title{Fundamental decoherence from quantum gravity: a pedagogical review}

\author{Rodolfo Gambini}
\address{Instituto de F\'{\i}sica, Facultad de Ciencias, 
Universidad
de la Rep\'ublica, Igu\'a 4225, CP 11400 Montevideo, Uruguay}
\author{Rafael A. Porto}
\address{Department of Physics, Carnegie Mellon University,
Pittsburgh, PA 15213}
\author{Jorge Pullin}
\address{Department of Physics and Astronomy, 
Louisiana State University, Baton Rouge,
LA 70803-4001}

\date{March 20th 2006}

\begin{abstract}
  We present a discussion of the fundamental loss of unitarity that
  appears in quantum mechanics due to the use of a physical apparatus
  to measure time.  This induces a decoherence effect that is
  independent of any interaction with the environment and appears in
  addition to any usual environmental decoherence.  The discussion is
  framed self consistently and aimed to general physicists. We derive
  the modified Schr\"odinger equation that arises in quantum mechanics
  with real clocks and discuss the theoretical and potential
  experimental implications of this process of decoherence.
\end{abstract}

\maketitle
\section{Introduction}

As ordinarily formulated, quantum mechanics involves and idealization.
The idealization is the use of a perfect classical clock to measure
times. Such a device clearly does not exist in nature, since all
measuring devices are subject to some level of quantum fluctuations.
Therefore the equations of quantum mechanics, when cast in terms of
the variable that is really measured by a clock in the laboratory,
will differ from the traditional Schr\"odinger description. Although
this is an idea that arises naturally in ordinary quantum mechanics,
it is of paramount importance when one is discussing quantum gravity.
This is due to the fact that general relativity is a generally
covariant theory where one needs to describe the evolution in a
relational way. One ends up describing how certain objects change when
other objects, taken as clocks, change. At the quantum level this
relational description will compare the outcomes of measurements of
quantum objects. Quantum gravity is expected to be of importance in
regimes (e.g. near the big bang or a black hole singularity) in which
the assumption of the presence of a classical clock is clearly
unrealistic. The question therefore arises: is the difference between
the idealized version of quantum mechanics and the real one just of
interest in situations when quantum gravity is predominant, or does it
have implications in other settings? We will argue that indeed
it does have wider implications.  Some of them
are relevant to conceptual questions (e.g.  the problem
of measurement in quantum mechanics or the black hole information
paradox) and there might even be experimental implications.

Although we have discussed several of these issues in previous papers
\cite{cqg,njp,piombino}, the latter were written with an audience of
relativists and quantum gravity experts in mind, and involving a
particular approach to quantum gravity we have been pioneering
\cite{prldiscrete,ashtekar}. Since many of the results are really robust and
independent of details of quantum gravity, we are giving a
presentation in this paper with minimal references to issues that may
be unfamiliar to physicists from outside the research area of
gravitational physics.

The plan of this paper is as follows: in the next section we will
derive the form of the evolution equation of quantum mechanics when
the time variable, used to describe it, is measured by a real clock.
In section III we will consider a fundamental bound on how accurate
can a real clock be and the implications it has for quantum mechanics
in terms of real clocks and its consequences. Section IV discusses the
implications  of the formalism.

\section{Quantum mechanics with real clocks}

Given a physical situation described by a (multi-dimensional) phase
space $q,p$, we start by choosing a ``clock''. By this we mean a
physical quantity (more precisely a set of quantities, like when one
chooses a clock and a calendar to monitor periods of more than a day)
that we will use to keep track of the passage of {\it time}. An
example of such a variable could be the angular position of the hand
of an analog watch. Let us denote it by $T(q,p)$. We then identify
some physical variables that we wish to study as a function of time.
We shall call them generically $O(q,p)$ (``observables''). We then
proceed to quantize the system by promoting all the observables and
the clock variable to self-adjoint quantum operators acting on a
Hilbert space. The latter is defined once a well defined inner
product is chosen in the set of all physically allowed states.
Usually it consists of squared integrable functions $\psi(q)$.

Notice that we are not in any way modifying quantum mechanics. We
assume that the system has an evolution in terms of an external
parameter $t$, which is a classical variable, given by a Hamiltonian and
with operators evolving with Heisenberg's equations (it is easier to
present things in the Heisenberg picture, though it is not mandatory
to use it for our construction). Then the standard rules of quantum 
mechanics and its probabilistic nature apply.

We will call the eigenvalues of the
``clock'' operator $T$ and the eigenvalues of the ``observables'' $O$.
We define the projector associated to the measurement of the time variable
within
the interval $[T_0-\Delta T,T_0+\Delta T]$,
\begin{equation}
P_{T_0}(t) =\int_{T_0-\Delta T}^{T_0+\Delta T} dT \sum_k |T,k,t><T,k,t|
\end{equation}
where $k$ denotes the eigenvalues of the operators that form a
complete set with $\hat{T}$ (the eigenvalues can have continuous or
discrete spectrum, in the former case the sum should be replaced by an
integral).  We have assumed a continuous spectrum for $T$ therefore
the need for the integral over an interval on the right hand side.
The interval $\Delta T$ is assumed to be very small compared to any of
the times intervals of interest in the problem, in particular the time
separating two successive measurements. Similarly we introduce a
projector associated with the measurement of the observable $O$,
\begin{equation}
P_{O_0}(t) =\int_{O_0-\Delta O}^{O_0+\Delta O} dO \sum_j |O,j,t><O,j,t|
\end{equation}
with $j$ the eigenvalues of a set of operators that form a complete
set with $\hat{O}$. These projectors have the usual properties, i.e.,
$P_a(t)^2 = P_a(t)$, $\sum_a P_a(t)=1, \forall t$ and
$P_a(t)P_{a'}(t)= 0$ if the intervals surrounding $a$ and $a'$ do not
have overlap.

We would like now to ask the question ``what is the probability that
the observable $O$ take a given value $O_0$ given that the clock
indicates a certain time $T_0$ ''. Such question is embodied in the
conditional probability,
\begin{equation}\label{condprob}
{\cal P}\left( O\in [O_0-\Delta O,O_0+\Delta O]|T \in 
[T_0-\Delta T,T_0+\Delta T]\right) =\lim_{\tau\to\infty}
{\int_{-\tau}^{\tau} dt\, {\rm Tr}\left(P_O(t) P_T(t)\rho P_T(t)\right) 
\over 
{\int_{-\tau}^{\tau} dt\,{\rm Tr}\left(P_T(t)\rho \right)}}
\end{equation}
where we have used the properties of both projectors and the integrals
over $t$ in the right hand side are taken over all its possible
values.  The reason for the integrals is that we do not know for what
value of the external ideal time $t$ the clock will take the value
$T_0$. In this expression $\rho$ is the density matrix of the system.
One has to take some obvious cares, like for instance to choose a
clock variable (or set of variables) that do not take twice the same
value during the relevant lifetime of the experiment one is
considering.

The above expression is general, it will apply to any choice of
``clock'' and ``system'' variables we make. The relational evolution
of the conditional probabilities will be complicated and will bear
little resemblance to the usual evolution of probabilities in ordinary
quantum mechanics unless we make a ``wise'' selection of the clock and
system variables. What we mean by this is that we would like to choose
as clock variables a subsystem that interacts little with the system
we want to study and that behaves semiclassically with small quantum
fluctuations. Namely, the physical clock will be correlated with the
ideal time in such a way to produce the usual notion of time. In
such a regime one expects to recover ordinary Schr\"odinger evolution
(plus small corrections) even if one is using a ``real'' clock. Let us
consider such a limit in detail. We will assume that we divide the
density matrix of the whole system into a product form between clock
and system, $\rho =\rho_{\rm cl}\otimes \rho_{\rm sys}$ and the
evolution will be given by a unitary operator also of product type
$U=U_{\rm cl}\otimes U_{\rm sys}$.

Up to now we have considered the quantum states as described by a 
density matrix at a time $t$. Since the latter is unobservable, we 
would like to shift to a description where we have density matrices
as functions of the observable time $T$. To do this, we recall the 
expression for the usual probability in the Schr\"odinger representation
of measuring the value $O$ at a time $t$,
\begin{equation}
  {\cal P}\left(O|t\right)\equiv {{\rm Tr}\left(P_{O}(0)\rho(t)\right)
\over {\rm Tr}\left(\rho(t)\right)}\label{ordinary}
\end{equation}
where the projector is evaluated at $t=0$ since in the Schr\"odinger
picture operators do not evolve. We would like to get a similar expression
in terms of the real clock. To do this we consider the conditional
probability (\ref{condprob}), and make explicit the separation between
clock and system,
\begin{eqnarray} \label{condprob2}
{\cal P}\left( O\in [O_0\pm\Delta O]|T \in 
[T_0\pm\Delta T]\right) &=&
\lim_{\tau\to\infty}
{\int_{-\tau}^{\tau} dt\,{\rm Tr}\left(U_{\rm sys}(t)^\dagger P_O(0)
U_{\rm sys}(t) U_{\rm cl}(t)^\dagger P_T(0) U_{\rm cl}(t)\rho_{\rm sys}\otimes 
\rho_{\rm cl}\right) 
\over 
{\int_{-\tau}^{\tau} dt\,{\rm Tr}\left(P_T(t)\rho_{\rm cl} \right) {\rm Tr}\left(\rho_{\rm sys}\right)}}\\
&=&
\lim_{\tau\to\infty}
{\int_{-\tau}^{\tau} dt\,{\rm Tr}\left(U_{\rm sys}(t)^\dagger P_O(0)
U_{\rm sys}(t) \rho_{\rm sys}\right){\rm Tr}\left(U_{\rm cl}(t)^\dagger P_T(0) U_{\rm cl}(t) 
\rho_{\rm cl}\right) 
\over 
{\int_{-\tau}^{\tau} dt\,
{\rm Tr}\left(P_T(t)\rho_{\rm cl} \right) {\rm Tr}\left(\rho_{\rm sys}\right)}}.
\end{eqnarray}

We define the probability that the resulting measurement of the clock
variable $T$ correspond to the value $t$,
\begin{equation}
{\cal P}_t(T) \equiv {{\rm Tr}\left(P_T(0) U_{\rm cl}(t)\rho_{\rm cl} U_{\rm cl}(t)^\dagger\right)\over
\int_{-\infty}^\infty dt\,{\rm Tr}\left(P_T(t) \rho_{\rm cl}\right)},
\end{equation}
and notice that $\int_{-\infty}^\infty dt {\cal P}_t(T)=1$. We now
define the evolution of the density matrix,
\begin{equation}
\rho(T) \equiv \int_{-\infty}^\infty U_{\rm sys}(t) \rho_{\rm sys} 
 U_{\rm sys}(t)^\dagger {\cal P}_t(T)
\end{equation}
where we dropped the ``sys'' subscript in the left hand side since
it is obvious we are ultimately interested in the density matrix of
the system under study, not that of the clock. Noting that
\begin{equation}
  {\rm Tr}\left(\rho(T)\right)=\int_{-\infty}^\infty dt\, {\cal P}_t(T) 
{\rm Tr}\left(\rho_{\rm sys}\right)={\rm Tr}\left(\rho_{\rm sys}\right),
\end{equation}
one can equate the conditional probability (\ref{condprob2}) to the
ordinary probability of quantum mechanics (\ref{ordinary}). We have
therefore ended with the standard probability expression with an
``effective'' density matrix in the Schr\"odinger picture given by 
$\rho(T)$. By its very definition, it is immediate to see that in
the resulting evolution unitarity is lost, since one ends up with
a density matrix that is a superposition of density matrices 
associated with different $t$'s and that each evolve unitarily 
according to ordinary quantum mechanics.

Now that we have identified what will play the role of a density
matrix in terms of a ``real clock'' evolution, we would like to see
what happens if we assume the ``real clock'' is behaving
semiclassically. To do this we assume that ${\cal P}_t(T) =f(T-T_{\rm
  max}(t))$, where $f$ is a function that decays very rapidly for
values of $T$ far from the maximum of the probability distribution
$T_{\rm max}$. To make the expressions as simple as possible, let
us assume that $T_{\rm max}(t)=t$, i.e. the peak of the probability
distribution is simply at $t$. More general dependences can of course
be considered, altering the formulas minimally (for a more complete 
treatment see \cite{njp}). We will also assume
that we can approximate $f$ reasonably well by a Dirac delta, 
namely,
\begin{equation}
f(T-t)=\delta(T-t)+ a(T)\delta'(T-t)+b(T)\delta''(T-t)+\ldots,
\end{equation}
where the first term has a unit coefficient so the integral of the probability
is unit and we assume $b(T)>0$ so it represents extra width with respect to
the Dirac delta.

We now consider the evolution of the density matrix,
\begin{equation}
\rho(T)=\int_{-\infty}^\infty dt\,\rho_{\rm sys}(t) {\cal P}_t(T)=
\int_{-\infty}^\infty dt\,\rho_{\rm sys}(t) f(T-t)
\end{equation}
and associating a Hamiltonian with the evolution operator $U(t)=\exp(iHt)$,
we get,
\begin{equation}
\rho(T)=\rho_{\rm sys}(T)+a(T) [H,\rho_{\rm sys}(T)]-b(T) [H,[H,\rho_{\rm sys}(t)]],
\end{equation}
and we notice that there would be terms involving further commutators if
we had kept further terms in the expansion of $f(T-t)$ in terms of the
Dirac deltas.

We can now consider the time derivative of this expression, and get,
\begin{equation}
{\partial \rho(T)\over \partial T} =i\left(-1+{\partial a(T)\over \partial T}\right) 
[H,\rho(T)] +\left(a(T)-{\partial b(T)\over \partial T}\right) [H,[H,\rho(T)]].
\end{equation}
If we had considered a symmetric distribution (it is natural to
consider such distributions since on average one does not expect an
effect that would lead systematically to values grater or smaller than
the mean value), we see that one would have obtained the traditional
evolution to leading order plus a corrective term,
\begin{equation}
{\partial \rho(T)\over \partial T} =i [\rho(T),H] +\sigma(T) [H,[H,\rho(T)]].
\end{equation}
and the extra term is dominated by the rate of change of the width of 
the distribution $\sigma(T)=\partial b(T)/\partial T$. 

An equation of this form has been considered in the context of
decoherence due to environmental effects, it is called the Lindblad
equation \cite{lindblad},
\begin{equation}
\frac{d}{dt}\rho=-i[H,\rho]-{\cal D}(\rho)\label{rho},
\end{equation}
with
\begin{equation} {\cal D}(\rho)=\sum_n[D_n,[D_n,\rho]], \;\;\;
D_n=D_n^{\dagger},\;\;\; [D_n,H]=0,
\end{equation}
and in our case there is only one $D_n$ that is non-vanishing and it
coincides with $H$. This is a desirable thing, since it implies that
conserved quantities are automatically preserved by the modified
evolution.  Other mechanisms of decoherence coming from a different
set of effects of quantum gravity have been criticized in the past
because they fail to conserve energy
\cite{hag}. It should be noted that Milburn arrived at a similar
equation as ours from different assumptions \cite{milburn}. Egusquiza,
Garay and Raya derived a similar expression from considering
imperfections in the clock due to thermal fluctuations
\cite{egusquiza}. It is to be noted that such effects will occur in
addition to the ones we discuss here.

What is the effect of the extra term? To study this, let us pretend for
a moment that $\sigma(T)$ is constant. That is, the distribution in the
clock variable has a width that grows linearly with time. In that case,
the evolution equation is exactly solvable. If we consider a system
with energy levels, the elements of the density matrix in the energy
eigenbasis is given by,
\begin{equation}
  \rho(T)_{nm} = \rho_{nm}(0) e^{-i\omega_{nm} T} e^{-\sigma \omega_{nm}^2 T}
\end{equation}
where $\omega_{nm}=\omega_n-\omega_m$ is the Bohr frequency
corresponding to the levels $n,m$. We therefore see that the
off-diagonal elements of the density matrix go to zero exponentially
at a rate governed by $\sigma$, i.e. by how badly the clock's
wavefunction spreads. It is clear that a pure state is eventually
transformed into a completely mixed state by this process.\

The origin of the lack of unitarity is the fact that definite
statistical predictions are only possible by repeating an experiment.
If one uses a real clock, which has thermal and quantum fluctuations,
each experimental run will correspond to a different value of the
evolution parameter. The statistical prediction will therefore
correspond to an average over several intervals, and therefore its
evolution cannot be unitary. 

In a real experiment, there will be decoherence in the system under
study due to interactions with the environment, that will be superposed
on the effect we discuss. Such interactions might be reduced by
cleverly setting up the experiment. The decoherence we are discussing
here however, is completely determined by the quality of the clock used.
It is clear that if one does experiments in quantum mechanics with 
poor clocks, pure states will evolve into mixed states very rapidly.
The effect we are discussing can therefore be magnified arbitrarily
simply by choosing a lousy clock. This effect has actually been observed
experimentally in the Rabi oscillations describing the exchange of
excitations between atoms and field \cite{Br}.

\section{Fundamental limits to realistic clocks}

We have established that when we study quantum mechanics with a
physical clock (a clock that includes quantum fluctuations), unitarity
is lost, conserved quantities are still preserved, and pure states
evolve into mixed states. The effects are more pronounced the worse
the clock is.  Which raises the question: is there a fundamental
limitation to how good a clock can be? This question was first
addressed by Salecker and Wigner \cite{wigner}. Their reasoning went
as follows: suppose we want to build the best clock we can. We start
by insulating it from any interactions with the environment.  An
elementary clock can be built by considering a photon bouncing between
two mirrors. The clock ``ticks'' every time the photon strikes one of
the mirrors. Such a clock, even completely isolated from any
environmental effects, develops errors. The reason for them is that by
the time the photon travels between the mirrors, the wavefunctions of
the mirrors spread. Therefore the time of arrival of the photon
develops an uncertainty.  Salecker and Wigner calculated the
uncertainty to be $\delta t \sim \sqrt{t/M}$ where $M$ is the mass of
the mirrors and $t$ is the time to be measured (we are using units
where $\hbar=c=1$ and therefore mass is measured in 1/second). The
longer the time measured the larger the error. The larger the
mass of the clock, the smaller the error.

So this tells us that one can build an arbitrarily accurate clock just
by increasing its mass. However, Ng and Van Damme \cite{ng} pointed
out that there is a limit to this. Basically, if one piles up enough
mass in a concentrated region of space one ends up with a black hole.
Some readers may ponder why do we need to consider a concentrated
region of space. The reason is that if we allow the clock to be more
massive by making it bigger, it also deteriorates its performance.
For instance, in the case of two mirrors and a photon, if one makes
the mirror big, there will be uncertainty in its position due to
elastic effects like sound waves traveling across it, which will
negate the effect of the additional mass (see the discussion in
\cite{ngotro} in response to \cite{baez}).
  
A black hole can be thought of as a clock (as we will see it turns
out to be the most accurate clock one can have).  It has normal modes of
vibration that have frequencies that are of the order of the light
travel time across the Schwarzschild radius of the black hole.  (It is
amusing to note that for a solar sized black hole the frequency is in
the kilohertz range, roughly similar to that of an ordinary bell). The
more mass in the black hole, the lower the frequency, and therefore
the worse its performance as a clock. This therefore creates a tension
with the argument of Salecker and Wigner, which required more mass to
increase the accuracy. This indicates that there actually is a ``sweet
spot'' in terms of the mass that minimizes the error. Given a time to
be measured, light traveling at that speed determines a distance, and
therefore a maximum mass one could fit into a volume determined by
that distance before one forms a black hole.  That is the optimal
mass.  Taking this into account one finds that the best accuracy one
can get in a clock is given by $\delta T \sim T_{\rm Planck}^{2/3}
T^{1/3}$ where $t_{\rm Planck}=10^{-44}s$ is Planck's time and $T$ is
the time interval to be measured. This is an interesting result. On
the one hand it is small enough for ordinary times that it will not
interfere with most known physics. On the other hand is barely big
enough that one might contemplate experimentally testing it, perhaps
in future years.

With this absolute limit on the accuracy of a clock we can quickly
work out an expression for the $\sigma(T)$ that we discussed in the
previous section \cite{prl,piombino}. It turns out to be $\sigma(T)=
\left({T_{\rm Planck}}\over{T_{\rm max}-T}\right)^{1/3}T_{\rm
  Planck}$. With this estimate of the absolute best accuracy of a
clock, we can work out again the evolution of the density matrix for a
physical system in the energy eigenbasis. One gets
\begin{equation}
  \rho(T)_{nm} = \rho_{nm}(0) 
e^{-i\omega_{nm} T} e^{-\omega_{nm}^2  T_{\rm Planck}^{4/3} T^{2/3}}.
\end{equation}

So we conclude that {\em any} physical system that we study in the lab
will suffer loss of quantum coherence at least at the rate given by
the formula above. This is a fundamental inescapable limit. A pure
state inevitably will become a mixed state due to the impossibility of
having a perfect classical clock in nature.

\section{Possible experimental implications}

Given the conclusions of the previous section, one can ask what are
the prospects for detecting the fundamental decoherence we propose. If
one would like to observe the effect in the lab one would require that
the decoherence manifest itself in times of the order of magnitude of
hours, perhaps days at best. That requires energy differences of the
order of $10^{10}eV$ in the Bohr frequencies of the system. Such
energy differences can only be achieved in ``Schr\"odinger cat'' type
experiments.  Among the best candidates today are Bose--Einstein
condensates, which can have $10^6$ atoms in coherent states. However,
it is clear that the technology is still not there to actually detect
these effects, although it could be possible in forthcoming years. 

A point that could be raised is that atomic clocks currently have an
accuracy that is less than a decade of orders of magnitude worse than
the absolute limit we derived in the previous section. Couldn't
improvements in atomic clock technology actually get better than our
supposed absolute limit? This seems unlikely. When one studies in
detail the most recent proposals to improve atomic clocks, they
require the use of entangled states \cite{atomic} that have to remain
coherent. Our effect would actually prevent the improvement of atomic
clocks beyond the absolute limit!

Another point to be emphasized is that our approach has been quite
naive in the sense that we have kept the discussion entirely in terms
of non-relativistic quantum mechanics with a unique time across space.
It is clear that in addition to the decoherence effect we discuss
here, there will also be decoherence spatially due to the fact that
one cannot have clocks perfectly synchronized across space and also
that there will be fundamental uncertainties in the determination of
spatial positions.  We have not studied this in detail yet, but it
appears that this type of decoherence could be even more promising 
from the point of view of experimental detection \cite{SiJa}.

\section{Conceptual implications}

The fact that pure states evolve naturally into mixed states has 
conceptual implications in at least three interesting areas of physics.
We will discuss them separately.

\subsection{The black hole information paradox}

The black hole information paradox appeared when Hawking
\cite{hawking} noted that when quantum effects are taken into account,
black holes emit radiation like a black body with a temperature
$T_{\rm BH}=\hbar/(8\pi k G M)$ where $M$ is the black hole mass, $k$ is 
Boltzmann's constant and $G$ is Newton's constant. As the black hole
radiates, it loses mass, and therefore its temperature increases. This
process continues until the black hole eventually evaporates
completely and the only thing left is outgoing purely thermal
radiation. Now, suppose one had started with a pure quantum state of
enough mass that it collapses into a black hole.  After the
evaporation process, one is left with a mixed state (the outgoing
purely thermal radiation). In ordinary quantum mechanics this presents
a problem, since pure states cannot evolve into mixed states.  (For
further discussion and references on the paradox see \cite{paradox}).

On the other hand, we have argued that due to the lack of perfectly
classical clocks, quantum mechanics really implies that pure states do
evolve into mixed states. The question is: could the effect be fast
enough to render the black hole information paradox effectively
unobservable? On one hand we have argued that our effect is small. But
it is also true that black holes usually take a very long time to
evaporate. Of course, a full calculation of the evaporation of a black
hole would require a detailed modelling including quantum effects of
gravity that no one is in a position of carrying out yet.  We have
done a naive estimate \cite{prl,piombino} of how our effect would take
place in the case of an evaporating black hole. To this aim we have
assumed the black hole is a system with energy levels (this is a
common assumption in many quantum gravity scenarios), and that most of
the Hawking radiation is coming from a transition between two dominant
energy levels separated by a characteristic frequency dependent on the
temperature as
\begin{equation}
\omega_{12}(t) = {1 \over \left(8 \pi\right)^2
t_P }\left(t_P\over T_{\rm max}-t\right)^{1/3}
\end{equation}
with $T_{\rm max}$ the lifetime of the black hole (how long it takes to
evaporate) and the subscript $12$ denotes that it is the transition
frequency between the two states of the system.  Although this model
sounds simple-minded it just underlies the robustness of the
calculation: it just needs that the black hole have discrete energy
levels characterized by a separation determined by the temperature of
the black hole.  It is general enough to be implemented either
assuming the Bekenstein spectrum of area or the spectrum stemming
from loop quantum gravity \cite{BaCaRo}.
We assume that we start with the black hole in a pure
state which is a superposition of different energy eigenstates
(there is no reason to assume that the black hole is  exactly in
an energy eigenstate, which would imply a stationary state with no
radiation being emitted; as soon as one takes into account the
broadening of lines due to interaction one has to consider a
superposition of states within the same broadened level with a
time dependent separation with a similar behavior).
Therefore the density matrix has off-diagonal
elements.

A detailed calculation \cite{piombino} for the evolution of the
density matrix shows that,
\begin{equation}
|\rho_{12}(T_{\rm max})| \sim |\rho_{12}(0)| \left({M_P \over M_{\rm
BH}}\right)^{2 \over 3}.
\end{equation}

For astrophysical sized black holes, where $M_{\rm BH}$ is of the
order of the mass of the Sun, this indicates that the off diagonal
elements are suppressed by the time of evaporation by $10^{-28}$,
rendering the information puzzle effectively unobservable. What
happens for smaller black holes? The effect is smaller. So can one
claim that there still is an information puzzle for smaller black
holes? This is debatable.  After all, we do expect decoherence from
other environmental effects to be considerably larger than the one we
are considering here.  If one makes the holes too small, then none of
these calculations apply, and in fact the traditional Hawking
evaporation is not an adequate description, since one has to take into
account full quantum gravity effects. So we can say that the paradox
is effectively eliminated for large black holes and we cannot say for
sure for smaller ones using this simplified analysis.

\subsection{The measurement problem in quantum mechanics}

A potential conceptual application of the fundamental decoherence that
we discussed that has not been exploited up to now is in connection
with the measurement problem in quantum mechanics.  The latter is
related to the fact that in ordinary quantum mechanics the measurement
apparatus is assumed to be always in an eigenstate after a measurement
has been performed.  The usual explanation \cite{Schlossauser} for
this is that there exists interaction with the environment. This
selects a preferred basis, i.e., a particular set of quasi-classical
states that commute, at least approximately, with the Hamiltonian
governing the system-environment interaction. Since the form of the
interaction Hamiltonians usually depends on familiar ``classical''
quantities, the preferred states will typically also correspond to the
small set of ``classical'' properties. Decoherence then quickly damps
superpositions between the localized preferred states when only the
system is considered. This is taken as an explanation of the
appearance to a local observer of a ``classical'' world of
determinate, ``objective'' (robust) properties.

The main problem with such a point of view is how is one to
interpret the local suppression of interference in spite of the
fact that the total state describing the system-environment
combination retains full coherence. One may raise the question 
whether retention of the full coherence could ever lead to 
empirical conflicts with the ascription of definite values to 
macroscopic systems. The usual point of view is that it would
be very difficult to reconstruct the off diagonal elements of
the density matrix in practical circumstances. However, at least
as a matter of principle, one could indeed reconstruct such
terms (the evolution of the whole system remains unitary
\cite{omnes}).

Our mechanism of fundamental decoherence could contribute to
the understanding of this issue. In the usual system-environment
interaction the off-diagonal terms of the density matrix oscillate
as a function of time. Since the environment is usually considered
to contain a very large number of degrees of freedom, the common
period of oscillation for the off-diagonal terms to recover 
non-vanishing values is very large, in many cases larger than
the life of the universe. This allows to consider the problem
solved in practical terms. When one adds in the effect we 
discussed, since it suppresses exponentially the off-diagonal
terms, one never has the possibility that the off-diagonal
terms will see their initial values restored, no matter how
long one waits. 

More generally, the environment-induced decoherence leads naturally to
a reduced density matrix for the quantum system plus the measurement
apparatus that is approximately diagonal and therefore very similar to
a statistical mixture of pure states corresponding to different
outcomes of the measurement.  That implies that a measurement of an
observable that only pertains to the system plus the measurement
device cannot discriminate between the total pure state and a mixed
state. However, as it was extensively discussed by d'Espagnat the
formal identity between the reduced density matrix and a mixed-state
density matrix is frequently misinterpreted as implying that the
system is in a mixed state. As the system is entangled with the
environment the total system is still described by a pure state and no
individual definite state or set of possible states may be attributed
to a portion of the total system. As it is well known, measurements on
the environment will always allow us, in principle, to distinguish
between the reduced and mixed state density matrices.

The combined effect of these two forms of decoherence could allow to
understand the physical transition from a reduced density matrix to a
mixed state. In fact, the precise unitary evolution of the total
system is broken by the clock-induced decoherence destroying the
correlations. What remains to be studied is whether this effect is
sufficiently fast to avoid any possibility of distinguishing, not only
for all practical purposes but also on theoretical basis, between
these two kinds of density matrices.

In a nutshell, this is how our effect might help understand the
measurement problem. In a future paper we will expand more on this and
on other issues related to the measurement problem.

\subsection{Quantum computing}

In quantum computing, when one performs operations one is evolving
quantum states. If one wishes the computers to perform faster, one
needs to expend extra energy to evolve the quantum states. Based on
this premise, Lloyd \cite{lloyd} presented a fundamental limitation to
how fast quantum computers can be. Using the Margolus--Levitin
\cite{male} theorem he notes that in order to perform a computation in
a time $\delta T$ one needs to expend at least an energy $E\geq \pi
\hbar/(2\Delta T)$.  As a consequence, a system with an average energy
$E$ can perform a maximum of $n=2E/(\pi\hbar)$ operations per second.
For an ``ultimate laptop'' (a computer of a volume of one liter and
one kilogram of weight) the limit turns out to be $10^{51}$ operations
per second.

Such results assume the evolution is unitary. When it is not, as we have
argued in this paper, erroneous computations are carried out. 
Since the rate of decoherence we discussed increases with increased
energy differences, the rate of erroneous computations increases the
faster one wishes to make the computer.

Can't one error correct? After all, one expects quantum computers to
have errors due to decoherence from environmental factors. One can
indeed error-correct. But there are limitations to how fast this 
can be done. At its most basic level error correction is achieved by
duplicating calculations and comparing results. This requires
spatial communication, which is limited by the speed of light. Our
point is that one cannot simply error correct one's way out of
the fundamental decoherence effects.

For instance, for a NOT gate our effect implies that after a time
$T\sim \pi/(2E)$ we will have,
\begin{equation}
|\psi_0><\psi_0|\rightarrow
(1-\epsilon)|\psi_1><\psi_1|+\epsilon
|\psi_0><\psi_0|,
\end{equation}
with $\epsilon=4 T^{4/3}_{\rm Planck} T^{2/3} E^2$. How does this effect
influence, for instance, the ``ultimate laptop''?

We have to distinguish a bit between serial and parallel computing. In
serial computing one achieves speed by increasing the energy in each
qubit. This enhances our decoherence effect and significantly affects
the performance. In a parallel machine one increases the speed by
operating simultaneously on many qubits with lower energies per qubit
therefore lowering the importance of the effect we introduced. For a
machine with $L$ qubits and a number of simultaneous operations $d_P$ one
gets,
\begin{equation}
n\le \left({1\over t_{\rm P}}\right)^{4/7} \left({cL \over
R}\right)^{3/7}d_p^{4/7} \sim 10^{47} {\rm op/s}\label{n},
\end{equation}
where the last estimate was obtained by taking the values of
parameters for the ``ultimate laptop'' (for more details see
\cite{qc}).

This is actually four orders of magnitude stronger than the bound that
Lloyd found. If one had chosen a serial machine, the bound would have
been tighter, $10^{42}$ operations per second.

We therefore see that although the effect we introduced is far
from being achievable in quantum computers built in the next few
years, it can limit the ultimate computing power of quantum computer.
This is quite remarkable, given that it is a limit obtained involving
gravity. Few people could have foreseen that gravity would play any
role in quantum computation.

\section{Discussion}

We have argued that the use of realistic clocks in quantum mechanics implies
that pure states evolve into mixed states. Another way of putting this
is that we are allowing quantum fluctuations in our clock. Similar ideas
have been considered by Bonifacio, with a different formulation \cite{Bo}. 
In quantum gravity and quantum cosmology 
it is natural to consider the clock to be part of the system under
study. This is what motivated our interest in these issues, but it is
clear that the core of the phenomenon can be described without references
to quantum gravity, and that is what we have attempted to do in this
presentation.

In the immediate future the most attractive possibility is to 
generalize these results to consider spatial decoherence due to the
lack of a universal clock across spatial points. This may open further
possibilities for experimentally observing these effects.

Even in the absence of a direct possibility of detecting these
effects, they can have important conceptual implications, as we have
illustrated with the black hole information puzzle, quantum computing
and the problem of measurement in quantum mechanics.

\section{Acknowledgements}

We wish to dedicate this paper to Octavio Obreg\'on. It is fitting to
have a paper intended to a broad audience given Octavio's diverse
contributions to physics.

This work was supported in part by grants NSF-PHY0244335,
DOE-ER-40682-143 and DEAC02-6CH03000, and by funds of the Horace
C. Hearne Jr. Institute for Theoretical Physics, PEDECIBA (Uruguay)
and CCT-LSU.


\begin{references}
\bibitem{cqg}
R. Gambini, R. Porto, J. Pullin,
Class.\ Quant.\ Grav.\  {\bf 21}, L51 (2004)
[arXiv:gr-qc/0305098].
\bibitem{njp} R. Gambini, R. Porto, J. Pullin,
New J.\ Phys.\  {\bf 6}, 45 (2004)
[arXiv:gr-qc/0402118].
\bibitem{piombino} R. Gambini, R. Porto, J. Pullin, 
  Braz.\ J.\ Phys.\  {\bf 35}, 266 (2005)
  [arXiv:gr-qc/0501027].
\bibitem{prldiscrete}
R.~Gambini and J.~Pullin,
Phys.\ Rev.\ Lett.\  {\bf 90}, 021301 (2003)
[arXiv:gr-qc/0206055];
\bibitem{ashtekar} R. Gambini, J. Pullin, ``Consistent discrete
space-time'', in ``100 years of Relativity Space-time structure: Einstein and beyond'', 
A. Ashtekar, World Scientific, Singapore (2006).
\bibitem{lindblad} G. Lindblad, Commun. Math. Phys. 48, 119 (1976).
\bibitem{hag} See for instance J. Ellis, J. Hagelin, D.V. Nanopoulos,
and M. Srednicki, Nucl. Phys. B241 (1984) 381; T. Banks, M.E. Peskin,
and L. Susskind, Nucl. Phys. B244 (1984) 125.
\bibitem{milburn} G. J. Milburn, Phys. Rev {\bf A44}, 5401 (1991).
\bibitem{egusquiza}
I. Egusquiza, L. Garay, J. Raya, Phys. Rev. {\bf
A59}, 3236 (1999) 
[arXiv:quant-ph/9811009].
\bibitem{Br} 
D. M. Meekhof, C. Monroe, B. E. King, W. M. Itano, and
D. J. Wineland, Phys. Rev. Lett {\bf 76}, 1796 (1996); M. Brune,
F. Schmidt-Kaler, A. Maali, J. Dreyer, E. Hagley, J. M. Raimond, and
S. Haroche, Phys. Rev. Lett. {\bf 76}, 1800 (1996); R. Bonifacio,
S. Olivares, P. Tombesi, D. Vitali Phys. Rev. {\bf A61}, 053802 (2000)
\bibitem{wigner}
E. Wigner, Rev. Mod. Phys. 29, 255 (1957).
\bibitem{ng}
Y.~J.~Ng and H.~van Dam,
Annals N.\ Y.\ Acad.\ Sci.\  {\bf 755}, 579 (1995) 
[arXiv:hep-th/9406110];
Mod.\ Phys.\ Lett.\ A {\bf 9}, 335 (1994).
\bibitem{ngotro} 
  Y.~J.~Ng and H.~van Dam,
  Class.\ Quant.\ Grav.\  {\bf 20}, 393 (2003)
  [arXiv:gr-qc/0209021].
\bibitem{baez}
J.~C.~Baez and S.~J.~Olson,
  Class.\ Quant.\ Grav.\  {\bf 19}, L121 (2002)
  [arXiv:gr-qc/0201030].
\bibitem{prl} 
R.~Gambini, R.~A.~Porto and J.~Pullin,
  Phys.\ Rev.\ Lett.\  {\bf 93}, 240401 (2004)
  [arXiv:hep-th/0406260].
\bibitem{atomic} See for instance A. Andre, A. Sorensen, M. Lukin,
  Phys. Rev. Lett {\bf 92}, 230801 (2004) [arXiv:quant-ph/0401130].
\bibitem{SiJa} C. Simon, D. Jaksch Phys. Rev. {\bf A70}, 052104 (2004). 
\bibitem{hawking} S. Hawking, Commun. Math. Phys. {\bf 43}, 199 (1975).
\bibitem{paradox} See for instance S. Giddings, L. Thorlacius, in
``Particle and nuclear astrophysics and cosmology in the next
millennium'', E. Kolb (editor), World Scientific, Singapore (1996)
[arXiv:astro-ph/9412046]; for more recent references see
S.~B.~Giddings and M.~Lippert, [arXiv:hep-th/0402073] and D.~Gottesman
and J.~Preskill, JHEP {\bf 0403}, 026 (2004) [arXiv:hep-th/0311269].
\bibitem{BaCaRo} For a discussion of both spectra see,
M.~Barreira, M.~Carfora and C.~Rovelli,
Gen.\ Rel.\ Grav.\  {\bf 28}, 1293 (1996)
[arXiv:gr-qc/9603064].
\bibitem{Schlossauser} M. Schlosshauer, Rev. Mod. Phys. {\bf 76}, 1267 (2004) [arXiv:quant-ph/0312059].
\bibitem{omnes} R. Omn\`es, ``The interpretation of quantum mechanics'', 
Princeton Series in Physics, Princeton, NJ (1994).
\bibitem{lloyd} S. Lloyd, Nature {\bf 406}, 1047 (2000).
\bibitem{male} N. Margolus, L. Levitin, Physica {\bf D120}, 188
(1998).
\bibitem{qc} R. Gambini, R. Porto, J. Pullin, 
``Fundamental gravitational limitations to quantum computing,''
  [arXiv:quant-ph/0507262].
\bibitem{Bo} R. Bonifacio, Nuo. Cim. {\bf D114}, 473 (1999).
\end{references}
\end{document}